\documentclass[12pt]{iopart}
\usepackage{iopams} 
\usepackage{mathptmx}
\usepackage[utf8]{inputenc} 
\usepackage{graphicx}
\usepackage{epstopdf}
\usepackage[numbers,comma,square,sort&compress,merge]{natbib} 
\usepackage{textcomp}
\usepackage{xspace}
\usepackage[innercaption,ragged]{sidecap}

\usepackage{url}
\usepackage{eurosym}
\usepackage{color,soul} 
\usepackage[colorinlistoftodos,prependcaption,textsize=footnotesize,textwidth=1.4cm]{todonotes}

\usepackage{hyperref}
\hypersetup{pdftitle={Correlated and in-situ electrical transmission electron microscopy studies and related membrane-chip fabrication}, colorlinks, citecolor=blue}

\newcommand{\degree}{$^{\circ}$\xspace}
\newcommand{\celsius}{$^{\circ}$C\xspace}

\begin{document}


\topical[Correlated and in-situ electrical TEM studies and related membrane-chip fabrication]{Correlated and in-situ electrical transmission electron microscopy studies and related membrane-chip fabrication}
\author{ Maria Spies$^1$\footnote{These authors have contributed equally.}, Zahra Sadre-Momtaz$^1$\footnotemark[\ddagger], Jonas L\"{a}hnemann$^{2}$\footnote{Present address: Paul-Drude-Institut für Festkörperelektronik, Leibniz-Institut im Forschungsverbund Berlin e.V., Hausvogteiplatz 5--7, 10117 Berlin, Germany}, Minh Anh Luong$^3$, Bruno Fernandez$^1$, Thierry Fournier$^1$, Eva Monroy$^2$, and Martien I. den Hertog$^{1}$}
\ead{martien.den-hertog@neel.cnrs.fr }
\address{$^1$ Univ. Grenoble-Alpes, CNRS, Institut N\'eel, 25 av.\ des Martyrs, 38000 Grenoble, France}
\address{$^2$ Univ. Grenoble-Alpes, CEA, IRIG-PHELIQS, 17 av.\ des Martyrs, 38000 Grenoble, France}
\address{$^3$ Univ. Grenoble-Alpes, CEA, IRIG-MEM, 17 av.\ des Martyrs, 38000 Grenoble, France}

\begin{abstract}

Understanding the interplay between the structure, composition and opto-electronic properties of semiconductor nano-objects requires combining transmission electron microscopy (TEM) based techniques with electrical and optical measurements on the very same specimen. Recent developments in TEM technologies allow not only the identification and in-situ electrical characterization of a particular object, but also the direct visualization of its modification in-situ by techniques such as Joule heating. Over the past years, we have carried out a number of studies in these fields that are reviewed in this contribution. In particular, we discuss here i) correlated studies where the same unique object is characterized electro-optically and by TEM, ii) in-situ Joule heating studies where a solid-state metal-semiconductor reaction is monitored in the TEM, and iii) in-situ biasing studies to better understand the electrical properties of contacted single nanowires. In addition, we provide detailed fabrication steps for the silicon nitride membrane-chips crucial to these correlated and in-situ measurements.
\end{abstract}



\submitto{\NT}

\ioptwocol

\maketitle

\section{Introduction}

In the current age of nano-fabrication, there is a general demand for characterization methods with a high spatial resolution. Transmission electron microscopy (TEM) today achieves a spatial resolution well below typical inter-atomic distances, allowing characterization of the crystal lattice with atomic resolution. TEM is therefore an immediate choice to study nano-materials like nanowires (NWs) or nanotubes with the aim to better understand the physical properties originating from their atomic structure and composition. It is common practice to perform TEM on one or few of such nano-objects, and correlate these data with results from other characterization techniques extracted from a different set of nano-objects that were nominally identical (fabricated/grown under the same conditions). However, a common problem in the fabrication/growth of nano-structures is that the presence of small differences between the objects synthesized on the same wafer are unavoidable with current state-of-the-art fabrication/growth methods, and such small differences can have a crucial impact on their properties. Therefore, it is important to perform correlated studies where all characterization techniques are applied to the very same object, in order to better understand its properties and facilitate modelling of exactly the structure under study.

For example, it has been shown by the correlation of TEM with electrical transport experiments that the conductive behaviour (metallic or semiconductor) of In$_2$Se$_3$ NWs depends on the crystallographic growth direction of the NWs \cite{peng2008large}. The effect of the hexagonal wurtzite or cubic zincblende crystal structure on the bandgap of GaAs NWs was elucidated by photoluminescence (PL) in the framework of correlated studies \cite{ahtapodov_story_2012,Heiss_2011,Vainorius_2014,Vainorius_2016}. For InP NWs, it was found that the PL of single crystalline NWs was very different from the PL of NWs containing twin defects \cite{InP_Pl}, while the thickness of cubic insertions in hexagonal GaP is directly linked to their emission energy \cite{Assali_nl_2017}. The effect of polytypism was also studied and is less pronounced in Ga(N)P NWs \cite{dobrovolsky_effects_2015}. The cathodoluminescence from (In,Ga)N insertions in GaN NWs was correlated to the location of the insertions observed in TEM \cite{Lahnemann_2011}. The above-mentioned studies were largely carried out by depositing the specimens on commercial silicon nitride membrane-chips or commercial carbon grids, which are then mounted on the TEM microscope sample holder.

It is interesting to perform the TEM visualisation of the specimen at the beginning and at the end of a specific treatment to understand its effect at atomic length scales. However, it is certainly more powerful to study the effect of the treatment in-situ or in-operando in TEM with atomic resolution, in order to precisely understand the rearrangement of atoms during the reaction process, e.g. while the specimen is subject to heat or electrical stress. Recently, significant advances in TEM sample holders have opened the possibility to implement a variety of in-situ experiments, including in-situ heating and biasing, as well as allowing measurements in liquid and gaseous environments. Therewith, it becomes possible to observe processes in-situ \cite{spiecker2019insights, den2018situ, el2019situ} or in-operando, rather than ‘just’ doing a pre- and post-mortem analysis.

Different approaches exist to perform electrical in-situ TEM. One possibility is to have a moveable tip on one side, and a static electrical contact on the other, to first make the electrical connection in-situ in the TEM, followed by the in-situ biasing \cite{he_silicon_2013,zheng_three-dimensional_2019}. While this approach is certainly interesting and may allow contacting a large variety of samples, the drawback of making one electrical contact in-situ with a tip can be that this contact is not necessarily of high quality, well controlled and reproducible.

A membrane-based chip with electrical contacts is an ideal platform for heating experiments, which can be easily combined with electrical biasing. If a suitable membrane-chip material with low thermal conductivity is chosen (e.g. silicon nitride \cite{chen2019situ} or ceramic-based membrane-chips), the electrical contacts allow controlling the sample temperature via Joule heating, by flowing a current through a metal line or spiral on the chip. Since only a very small volume is heated, the injected power remains minor, and therefore the drift of the sample and TEM holder upon heating is small. 
This configuration is a major improvement compared with more traditional heating holders, where the entire tip of the holder is heated, resulting in thermal gradients in the holder and requiring long stabilization times (minutes). 
While the commercial solutions that can be acquired for heating (and/or biasing) are certainly interesting, we have developed a more flexible approach to fabricate custom membrane-chips optimised for correlated studies and electrical biasing experiments in the TEM. 

We have performed numerous studies using these membrane-chips focusing on semiconducting NWs of different materials such as Si, Ge, GaN/AlN and ZnO. 
The aim of this paper is to demonstrate the potential this approach by giving an overview of the studies performed using this membrane-chip technology. These studies can be grouped into three categories: 
i) Correlated studies where the same unique object is characterized electro-optically ex-situ and by TEM -- either to relate specific luminescence features to particular structural defects or to understand the performance of single NW devices such as photodetectors or quantum-dot light emitters. 
ii) In-situ visualization of chemical diffusion processes using Joule heating applied to metal-semiconductor reactions in a NW. 
iii) In-situ bias-dependent TEM characterization using off-axis electron holography to assess doping concentration and depletion length at Schottky contacts on single NWs. For reference, the fabrication process of the membrane-chips is first described in detail.

\begin{figure*}[t]
\includegraphics[width=0.99\textwidth]{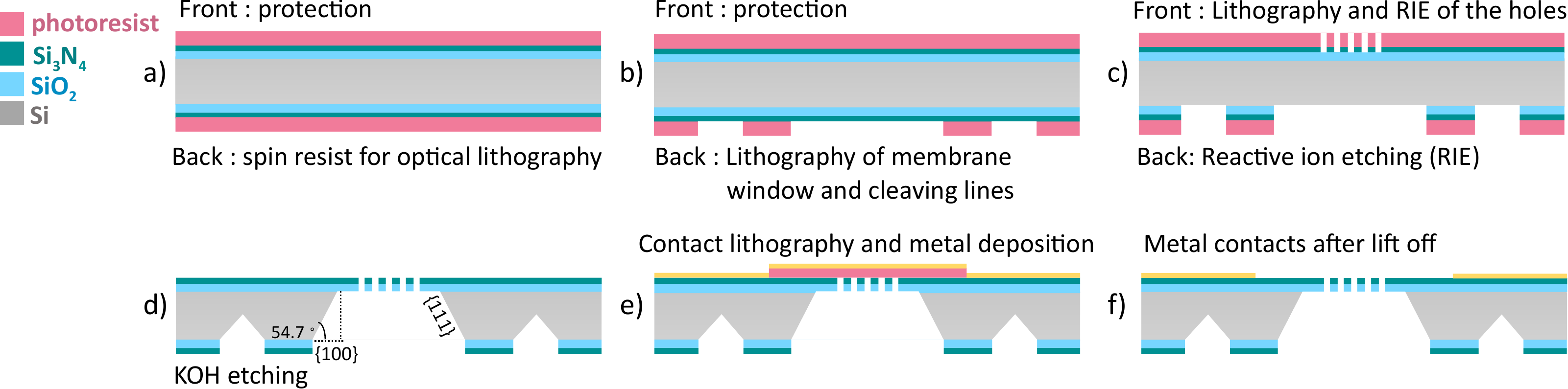}
\caption{\label{fig:CBdiamonds} Membrane-chip fabrication process: (a) Spinning photo-resist on both sides of a 4 inch wafer: on the front side in order to protect it from being scratched and on the back side as a preparation for the laser lithography step. (b) First laser lithography on the back side to introduce membrane area and cleave lines. (c) RIE on the back side to remove the Si$_\mathbf{3}$N$_\mathbf{4}$ and SiO$_\mathbf{2}$ layers  followed by the second laser lithography step on the wafer front side to introduce slits. (d) KOH bath to etch 400~\textmu{}m Si layer from the backside at the surfaces not protected by Si$_\mathbf{3}$N$_\mathbf{4}$. (e) Third laser lithography exposure step on the front side to define the contact pads and EBL markers on the membrane-chip followed by metal evaporation for contacts and (f) lift-off procedure.}
\end{figure*}

\section{Membrane-chip fabrication}

We have developed an approach to fabricate contacted nanostructures on a silicon nitride membrane-chip suitable for in-situ TEM analysis. 

We produce full four-inch wafers containing 384 membrane-chips, ordered in $2\times2$ and $4\times4$ arrays. 
Therefore, contacting of nano-objects can be done on such an array of membrane-chips, which allows to perform the process on 
several chips, each with several nano-objects, all in one run. 

Simultaneous processing of several chips facilitates the lithography steps and significantly reduces the fabrication time per chip.
The chip design can be adapted to the different TEM sample holders required for various experiments (TEM, micro-PL (\textmu{}-PL), energy dispersive X-ray spectroscopy (EDX), etc.). This includes making choices on the chip size, membrane size and thickness, nature of  the contacts, presence of metal markers and calibration fields for electron beam lithography (EBL). 

The chip fabrication process starts with 400~\textmu{}m thick, highly \emph{n}-doped (As$^{++}$) Si(100) four inch wafers. 
The thickness of the wafer is primarily chosen to be compatible with the TEM sample holder that will be used. On both sides, the wafer is covered by 200~nm of thermal SiO$_{2}$ and 40 or 200~nm stoichiometric Si$_{3}$N$_{4}$ synthesized by low-pressure chemical vapour deposition by LioniX International.
The thermal SiO$_{2}$ layer increases the electrical insulation and reduces the capacitance between the substrate and the metal pads deposited on top of the Si$_{3}$N$_{4}$ layer.
The choice of the Si$_{3}$N$_{4}$ thickness is a compromise to obtain high-resolution TEM images while keeping a robust membrane. It can be adapted either by changing the deposited layer thickness or by etching the Si$_{3}$N$_{4}$ layer using HF. 
For high-resolution scanning transmission electron microscopy (STEM), we typically use a Si$_{3}$N$_{4}$ thickness of 40~nm and a membrane surface of $100\times100$~\textmu{}m$^2$. For TEM techniques requiring the specimen to be suspended in vacuum (e.g. holography), we fabricate membranes including slit-shaped openings. In order to maintain membrane stability despite the openings, the Si$_{3}$N$_{4}$ layer thickness is increased to 200~nm, and the membrane surface is increased to $220\times220$~\textmu{}m$^2$.

The  membrane-chip fabrication steps are illustrated in figure~\ref{fig:CBdiamonds} for the case of a membrane with slits. For planar membranes without slits, the fabrication step of the slits is simply omitted. 

\begin{figure*}[t]
\includegraphics[width=0.99\textwidth]{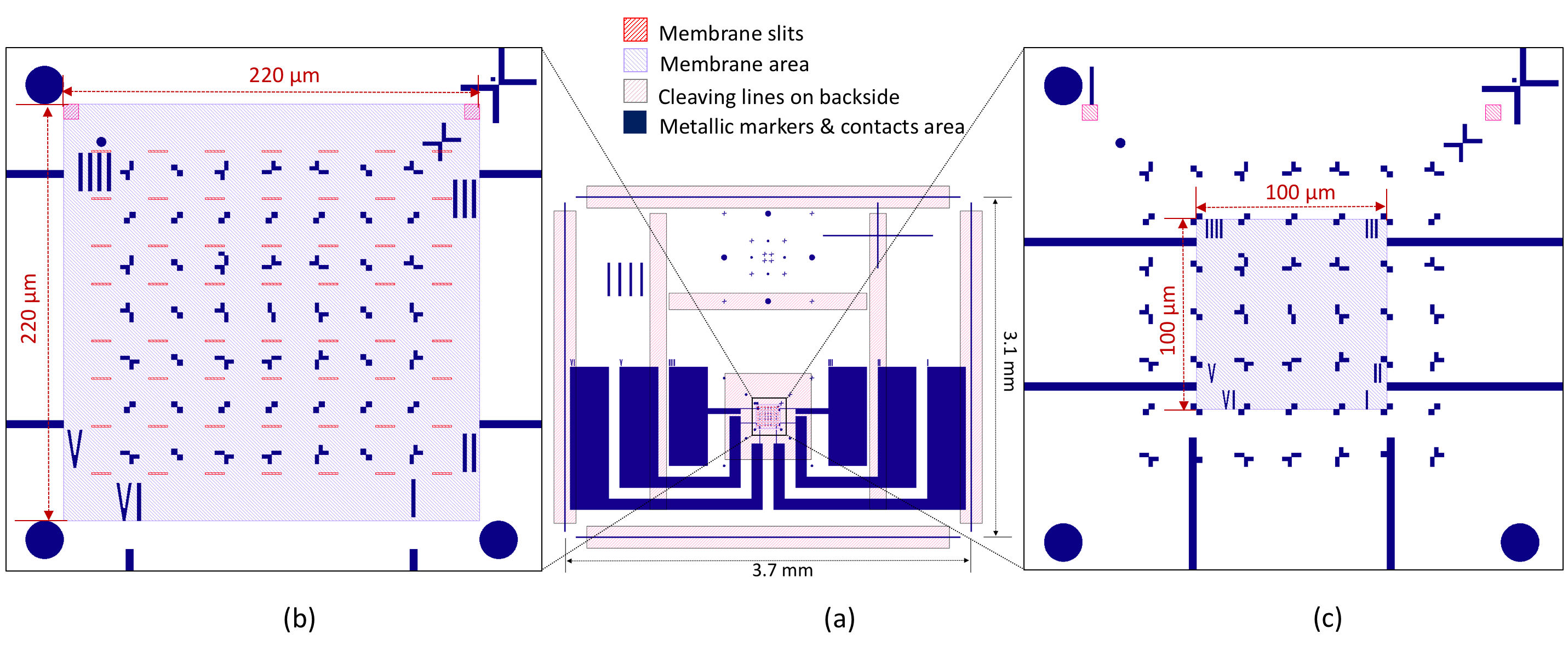}
\caption{\label{fig:CAD} (a) The membrane window, cleave lines, contact areas, labelling of the chips, labelling of the contact leads on the membrane and markers for EBL with all relevant dimensions of (b) membrane with slits and (c) planar membrane. }
\end{figure*}

Figure~\ref{fig:CAD} represents the drawings for the fabrication of both planar membranes and membranes with slits. Single chip dimensions of around $3.1 \times 3.7$ mm$^2$ are chosen, to fit a DENSsolutions double tilt six contacts heating/biasing TEM sample holder. The top part of the chip serves as a space where the chip can be handled with tweezers or clamped mechanically. The membrane itself is below the center of the chip and laterally centred. Three contact pads of around $0.3 \times 1.3$~mm$^{2}$ are placed on both sides of the membrane, each having a contact lead which connects to the membrane. On the membrane itself, markers are drawn as common reference for TEM or \textmu{}-PL experiments. As can be observed in Figure~\ref{fig:CAD}(b), additional cleave lines are added inside the chip region. Their width is smaller than the main cleave lines (150 \textmu{}m where the cleave lines between chips are 200 \textmu{}m) and they can be used to resize the chip once all characterization steps requiring electrical contacts are finished, to allow insertion in a standard TEM sample holder requiring samples fitting a ring of 3 mm diameter.

The first step in the fabrication process is to protect the front/sample side of the wafer by spinning a layer of photoresist (S1818, 30~s at 6000~r.p.m., accel.~4000 r.s$^{-2}$ and hotplate baked for 1~min at 115~\celsius) [figure~\ref{fig:CBdiamonds}(a)]. 

Subsequently, we spin the same photoresist on the other side of the wafer (back side) using the same parameters. 

\begin{table}[t]
    \caption{\label{tab:table1}RIE parameters used for etching Si$_{3}$N$_{4}$ and SiO$_{2}$ layers using SF$_{6}$ and CHF$_{3}$, respectively.}
      \begin{indented}
    \item[]\begin{tabular}{@{}llll} 
    \br
      \textbf{Etchant gas } & \textbf{Power} & \textbf{V$_{\textrm{bias}}$}& \textbf{Pressure}\\
       & (W) & (V) & (sccm) \\
    \mr
      SF$_6$ & 50 & 270 & 20\\
      CHF$_3$ & 50 & 440 & 15\\
    \br
    \end{tabular}
  \end{indented}
\end{table}

Then the membrane area and cleave lines are defined by UV laser lithography (Heidelberg Instruments DWL 66FS system with a 405--410~nm diode laser using 4.5\% of the full diode power of 120~mW) on the backside of the wafer. 

After development (1:1 solution of Microdev and de-ionized water (DI-H$_2$O), 1 minute) [figure~\ref{fig:CBdiamonds}(b)] the Si$_{3}$N$_{4}$ and SiO$_{2}$ layers are etched through opened windows in the photoresist using reactive ion etching (RIE) with SF$_{6}$ and CHF$_{3}$ in a Plassys RIE system with the parameters summarized in Table 1. 
The wafer is then cleaned in acetone and isopropyl alcohol (IPA) to remove the photoresist and potential contamination.

At this point, if the membrane should include slits, they are patterned by laser lithography on the front side of the wafer (figures~\ref{fig:CAD}(b) and \ref{fig:Slits on membrane}) followed by the Si$_{3}$N$_{4}$ etching step.


The laser lithography system employed includes a camera mounted below the sample stage that allows alignment of the wafer using features on its backside. It is important to minimise the rotation of the wafer with respect to the drawing below 3~mrad.



The wafer is then immersed in a potassium hydroxide (KOH) bath to etch the Si where it is exposed. A KOH solution is prepared by dilution of 756~g of KOH ($\geq$ 85\%, Ph.Eur., pure pellets) in 950~ml DI-H$_2$O. This solution is maintained at 80~\celsius. 
It is very helpful to add a dummy sample into the same KOH bath in order to determine the precise etching rate of Si of the prepared solution---typically around 70~\textmu{}m/h. Around one extra hour should be allocated for etching the SiO$_{2}$ of the front side. 

\begin{figure}[t]
\includegraphics[width=0.49\textwidth]{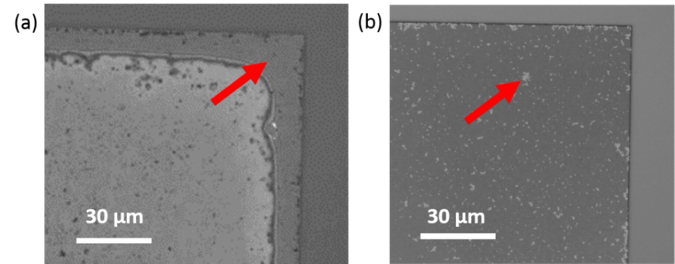}
\caption{\label{fig:koh} Visible light microscope images of the top right quadrant of a membrane with etching problems: (a) A rim of remaining SiO$_{2}$ on the edges of the membrane can be seen. (Careful not to mistake these for remaining solvents in the process of drying.) (b) Remaining SiO$_{2}$ particles on the backside of the membrane.}
\end{figure}
\begin{figure}[t]
\includegraphics[width=0.49\textwidth]{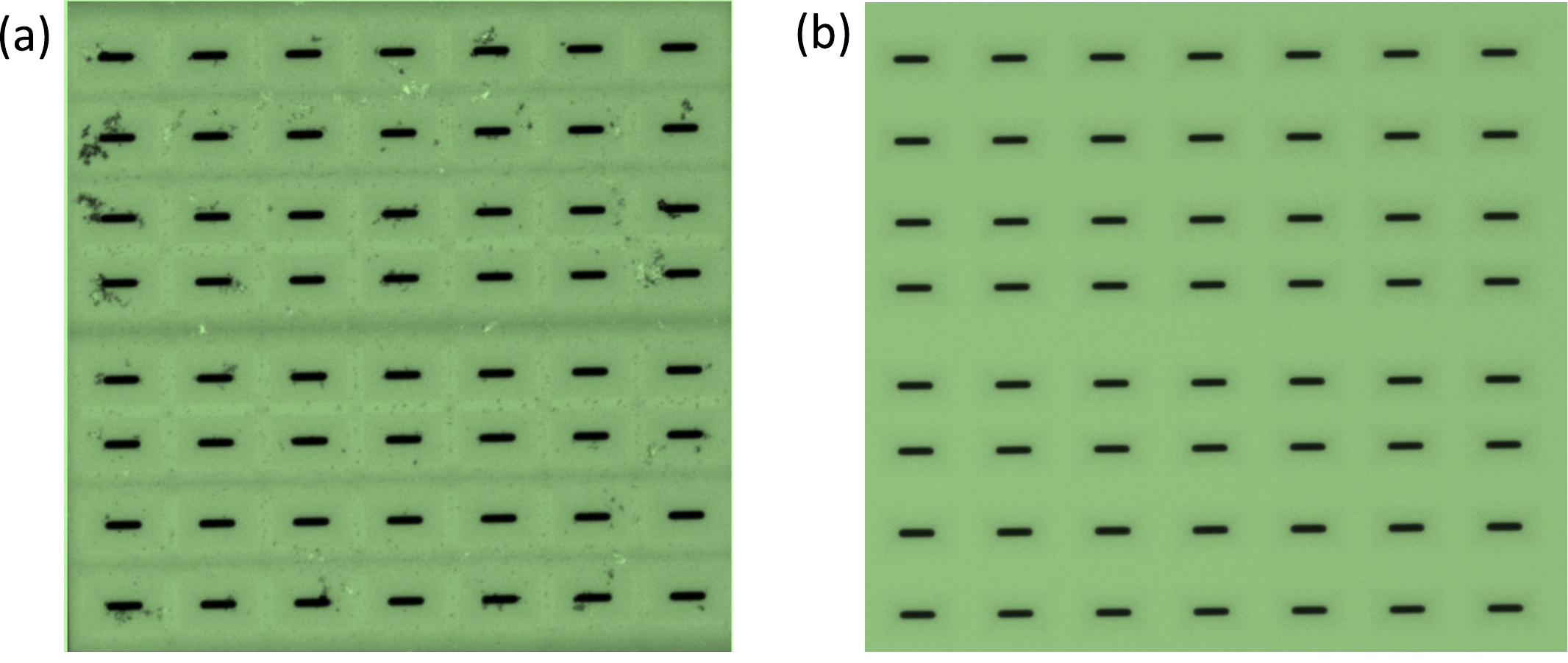}
\caption{\label{fig:Slits on membrane} Visible light microscope images showing the evolution of Si/SiO$_{2}$ etching on a membrane as a function of immersion time in a KOH bath: (a) Under-etched sample due to insufficient etching time in KOH (7.5~hours). The features around the holes are due to remaining SiO$_{2}$. (b) Fully etched membrane after longer etching time in KOH bath (9~hours).}
\end{figure}

In figure~\ref{fig:koh}, the result of insufficiently long KOH etch durations can be seen as observed by visible light microscopy on a planar membrane. Figure~\ref{fig:Slits on membrane} demonstrates the evolution of Si/SiO$_{2}$ etching of a membrane featuring slits for different etching times in the KOH bath. The granular structures around the slits in figure~\ref{fig:Slits on membrane}(a) represent incomplete etching of SiO$_{2}$ due to insufficient time in the KOH bath. By giving the sample more time in the bath, we finally end up with a clean, uniform membrane as shown in figure~\ref{fig:Slits on membrane}(b).

Since KOH etches Si along $\{111\}$ planes, the resulting angle with respect to the $\{100\}$ plane of the surface of the wafer is 54.7\degree [see figure~\ref{fig:CBdiamonds}(d)]. Therefore, the size of the membrane we obtain depends both on the size of the region that is opened and on the thickness of the wafer. For a wafer thickness of 400~\textmu{}m, we define squares of 666 (786)~\textmu{}m$^{2}$ to obtain windows of 100 (220)~\textmu{}m$^{2}$. Also the cleaving marks will result in triangular grooves with a depth of around one third of the wafer thickness. It is hence important to maintain a minimum separation of 20 \textmu{}m between different cleave lines.

After the KOH bath, the wafer is extensively rinsed with DI-H$_2$O and cleaned for 1 hour in HNO$_3$ (65\%) at 80~\celsius. 

\begin{figure*}[t]
\includegraphics[width=1\textwidth]{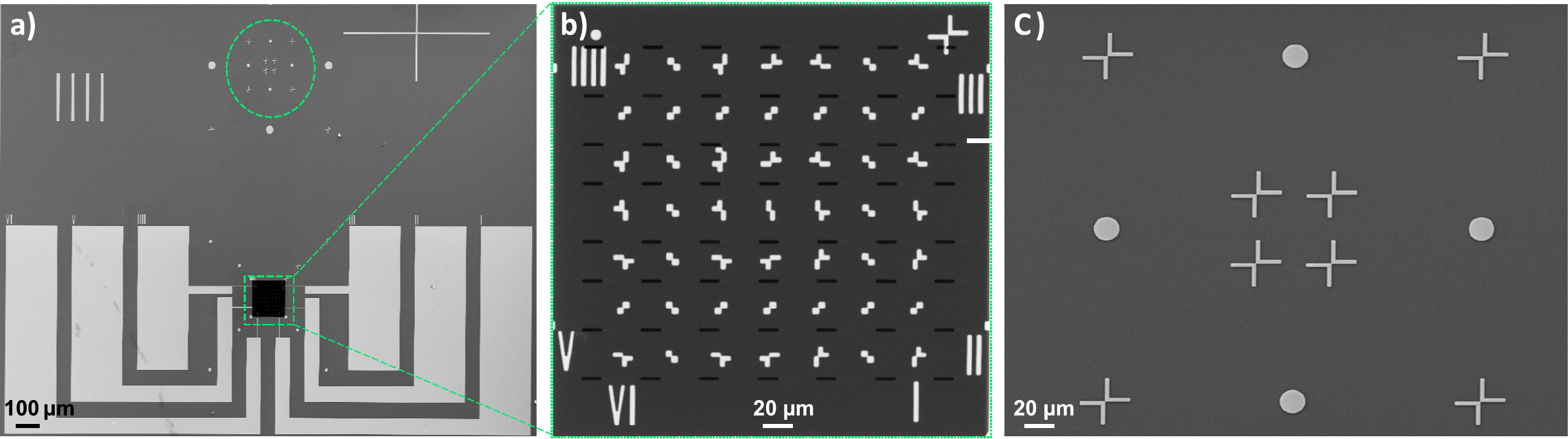}
\caption{\label{fig:chip+zoomed} Scanning electron micrographs of (a) a single membrane-chip with chip number, six contact pads leading to the membrane region and alignment markers. (b) Zoomed image of the membrane area with slits, and (c) Zoomed image of calibration field markers for EBL, as indicated by a dashed circle in (a).}
\end{figure*}

Finally, a lithography step on the front side of the wafer defines the metal pads and any desired markers or labels (Ti/Au, 5 nm / 35 nm, or Ti/Pt, 5 nm / 35 nm). Prior to metal deposition by electron beam evaporation, the sample is cleaned during 10~s using Ar plasma. 


To prepare the sample for the final laser lithography using 2.25\% of the laser power, LOR~3A and a positive photoresist (S1805) are spun on the wafer  (LOR~3A at 6000~r.p.m., accel. 4000 r.s$^{-2}$, 30~s, baked for 2~min at 200~\celsius, followed by S1805 at 6000~r.p.m., accel. 4000~r.s$^{-2}$, 30~s, baked for 1~min at 115~\celsius). 
The sample is developed in MF26A for 1~min and then rinsed with DI-H$_2$O.
%
After metal deposition
, the lift-off of the metal is done by putting the wafer in PG remover (based on N-methyl-2-pyrrolidone) at 80~\celsius for at least 2~hours. 
%
When the wafer is taken out, it is cleaned with IPA, rinsed with acetone and then put into an acetone bath. A pipette can be employed to agitate the acetone and help lift off the metal, which should proceed relatively easily. 
After rinsing the wafer with DI-H$_2$O and carefully blow drying with a standard nitrogen gun (low pressure, low angle with membrane-chip surface) the membrane-chips are ready to use. 

The scanning electron micrographs of a finished chip along with zoomed images of membrane area and calibration field markers are shown in figure~\ref{fig:chip+zoomed}(a)--(c). Several details of the metal layer, which can facilitate later experiments, can be observed in figure~\ref{fig:chip+zoomed}(a)--(c):

-	At and around the membrane, we define markers both for automatic and manual alignment, see figure~\ref{fig:chip+zoomed}(b).

-	On each chip, we define calibration fields of different sizes away from the membrane area to facilitate EBL, as shown in figure~\ref{fig:chip+zoomed}(c).

-	Each chip of an array is numbered for easy distinction. This number is given in large in the top left corner (visible in light microscopy), and is repeated in small close to the membrane (useful for example in optical spectroscopy experiments like \textmu{}-PL).

-	Each electrical contact has a number and this number is repeated at the end of the contact line on the membrane, so that it is visible during TEM observation.

-	The cleave lines on the backside are indicated by metal lines on the front side. Chips can easily be separated from one another without damaging the membrane by pressing on the front side at the cleaving line with a diamond tip. It is best to perform this step on a thick piece of cleanroom paper. 

\section{Review of studies based on custom membrane-chips}

These custom-made membrane-chips allow the realisation of a variety of correlated and in-situ microscopy experiments. We present here an overview, where we have differentiated:
\begin{itemize}
\item{Ex-situ studies correlating the opto-electrical properties of a single NW with its structural properties such as the crystal structure, composition and dimensions that are measured by TEM-based characterisation.} 
\item{In-situ Joule heating studies on a solid state metal-semiconductor reaction induced by heating where the evolution of the reaction is  directly visualised by TEM.}
\item{In-situ electrical biasing experiments where the variation of holographic measurements under bias can be used to analyse the behaviour of Schottky contacts and extract dopant concentrations.}
\end{itemize}

For all these experiments, as grown NWs were transferred from the growth substrate to the fabricated membrane-chips using a wet dispersion method, where a small piece of growth substrate was immersed in ethanol or iso-propanol. Using an ultrasonic bath, the NWs were detached from the substrate, and pipetting of the solution on the membrane-chips ensures a random dispersion of NWs. Then, after finding suitably nice and long single NWs using scanning electron microscopy (SEM), and localising their position with respect to metal markers, the NWs could be contacted by a sequential EBL step followed by metal evaporation or sputtering.

For most correlated and in-situ Joule heating experiments scanning TEM was used, carried out either on a a probe corrected FEI Titan, Titan Themis or Titan Ultimate also equipped with an image corrector. Some in-situ Joule heating experiments were carried out on a CM300. In-situ electrical biasing combined with off-axis electron holography was carried out on the Titan Ultimate. Some of these experiments were carried out at lower acceleration voltage (80 or 100 kV) to avoid modifying the electro-optical properties of the NW. Systematically, (S)TEM characterization was carried out after performing electrical or optical characterization.

\subsection{Ex-situ correlated microscopy studies}

\subsubsection{Optical signature of crystalline defects.}

\begin{figure}[t]
\includegraphics[width=0.49\textwidth]{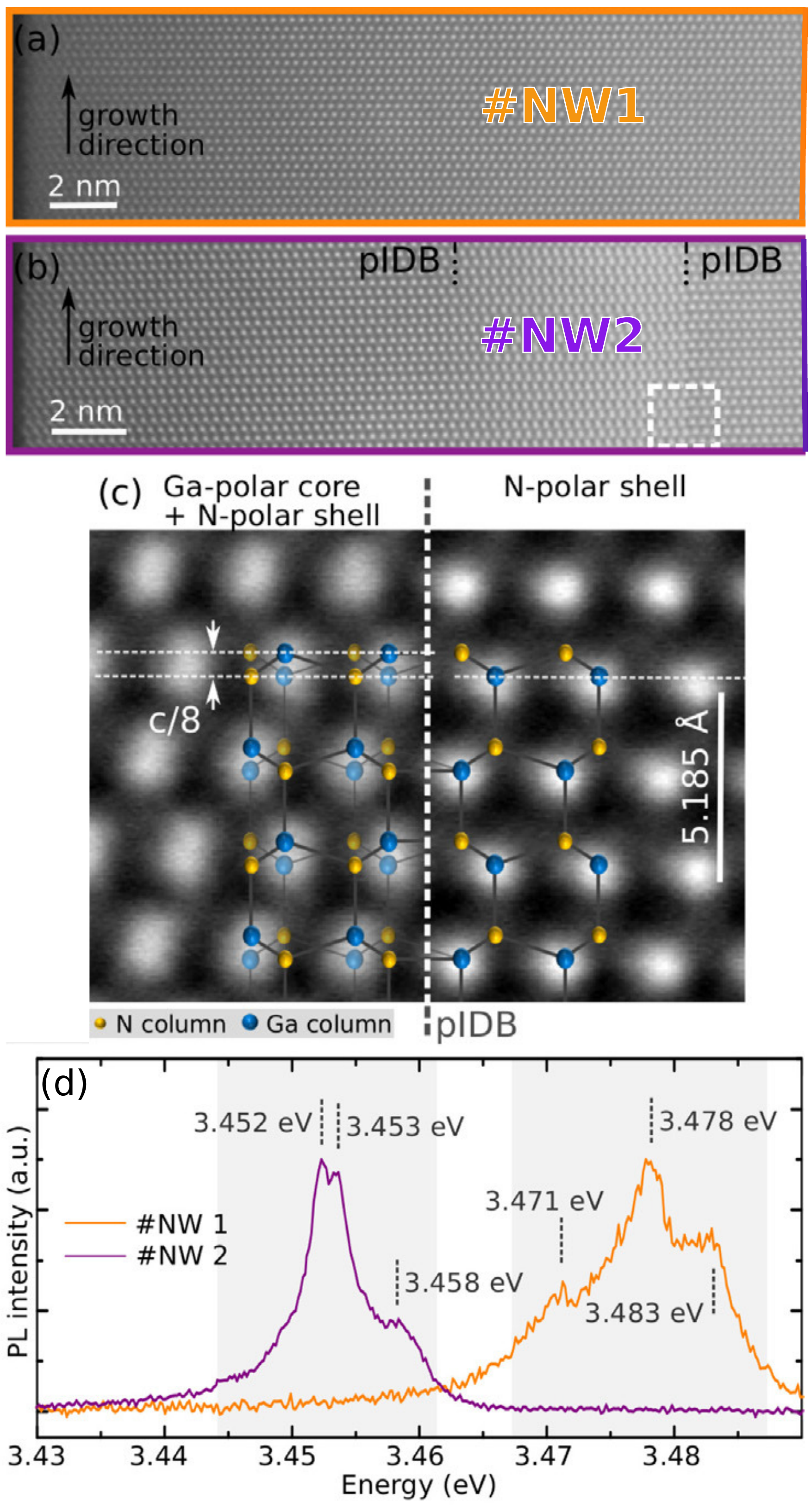}
\caption{\label{fig:auzelle_apl}(a) High-resolution HAADF STEM image of \#NW~1, which is free of inversion domains. (b) \& (c) High resolution HAADF STEM images of \#NW~2, which hosts a Ga-polar inversion domain. 
(d) Comparative \textmu-PL spectra of \#NW~1 and \#NW~2, acquired at 10~K. Reproduced from reference~\cite{auzelle_attribution_2015} with permission from AIP.}
\end{figure}

Correlated microscopy studies that link a crystalline feature and opto-electronic property are of utmost interest for device development.
They reveal the influence of structural disorder on, for example, the light emission properties of a nano-object. We present examples where the above-described membrane-chips were used in two studies relating specific defects in nanowires to their optical properties \cite{nogues_cathodoluminescence_2014,auzelle_attribution_2015}. The metal markers on the membrane-chip were used to identify and measure the same object in the different experiments, including TEM.

With regard to the optical effect of stacking faults in GaN, some reports had shown that 
I1 basal stacking faults could act as radiative recombination
centres in GaN \cite{Salviati_1999,Liu_2005,corfdir_2009}, whereas combined TEM and cathodoluminescence (CL) studies of GaN epilayers 
had demonstrated a correlation between the presence
of different types of stacking faults and well-identified emission
lines \cite{Salviati_1999,Liu_2005}. However, the density of structural defects was relatively
high in those samples.

To clarify this issue, Nogues \emph{et al.}\ \cite{nogues_cathodoluminescence_2014} carried out a correlated study of individual GaN NWs in SEM combined with low temperature cathodoluminescence, \textmu-PL, and STEM. 
The use of nanowires as material platform to explore the properties of stacking faults in GaN is motivated by the high crystalline quality of such nano-objects, where single stacking fault appear isolated from other crystalline defects. 
Nogues \emph{et al.}\ showed that NWs exhibiting well-localized regions emitting light at 3.42~eV presented a single stacking fault in these regions. This allowed the assignment of this luminescence line to a stacking fault bound exciton, and precise measurements of the CL signal intensity in the vicinity of the stacking fault gave access to the exciton diffusion length in this region.

Another feature commonly observed in the low-temperature PL spectrum of GaN NWs is a sub-bandgap 
transition at 3.45~eV \cite{calleja_2000,huang_origin_2015}. This emission had been attributed to two-electron-satellite
(TES) excitonic recombinations on a near-surface donor \cite{corfdir_2009_jap}
and to near-surface
point defects \cite{Brandt_2010,Furtmayr_2008}, respectively. However, both hypotheses were not coherent with magneto-luminescence
experiments and a polarization-resolved study \cite{Sam_2013}.
Using correlated high-resolution high angle annular dark field (HAADF) STEM [figure~\ref{fig:auzelle_apl}(a-c)] and \textmu-PL [figure~\ref{fig:auzelle_apl}(d)] on the same single NWs grown by plasma assisted molecular beam epitaxy (PA-MBE)
 the 3.45~eV luminescence of GaN NWs was unambiguously attributed to the presence of prismatic inversion domain boundaries in the  study presented by Auzelle and Haas \emph{et al.}\ \cite{auzelle_attribution_2015}.

\subsubsection{Heterostructured single nanowire devices: photodetectors.}

%

\begin{SCfigure*}[10][t]
\includegraphics[width=0.49\textwidth]{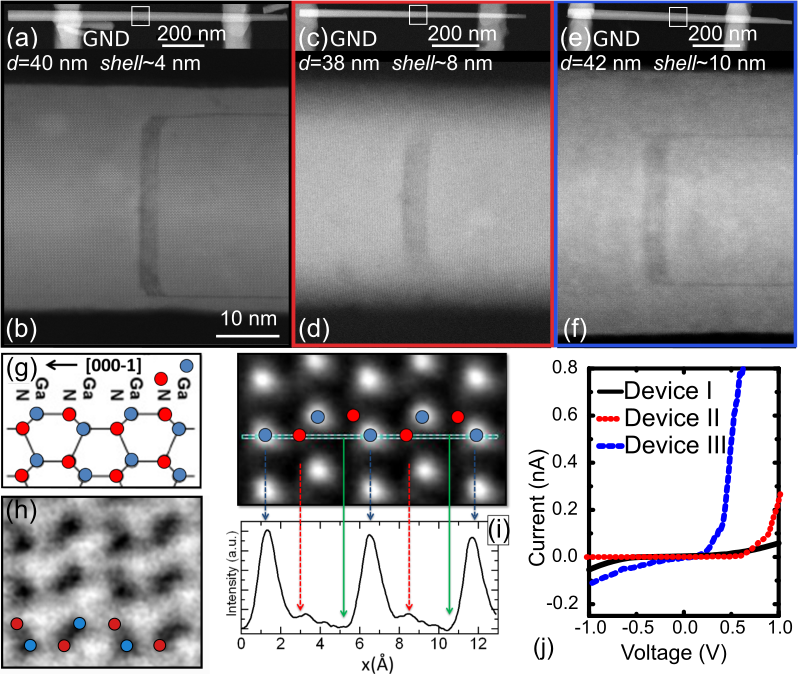}
\caption{\label{fig:hertog_nl} 
HAADF STEM images (a,c,e) and corresponding zoom of the area around the AlN barrier (b,d,f) of NW devices I, II, and III, respectively. Devices I and III are viewed along $[2\bar{1}\bar{1}0]$ 
while device II is viewed along $[10\bar{1}0]$. 
The diameter $d$ at the AlN barrier and the respective average GaN shell thickness ($shell$) are indicated in (b,d,f).  
(g) Schematic representation of the GaN crystal structure viewed along $[11\bar{2}0]$. 
(h) Convoluted atomic resolution ABF STEM and (i) averaged HAADF STEM image obtained on the NW shown in (a,b) viewed along the $[11\bar{2}0]$ axis with a superposition of the GaN atomic structure, combined with an intensity profile along the superimposed line. 
(j) Smoothed \emph{I}-\emph{V} characteristics measured in the dark. The top of the NW (left side in images a, c, and e) was connected to ground as indicated. Reproduced from reference~\cite{den_hertog_correlation_2012} with permission from ACS.}
\end{SCfigure*}

Research on NWs as photodetectors \cite{soci_nanowire_2010,vj_perspective_2011,spies_nanowire_2019} is motivated not only by miniaturisation issues, but also by the possibility to enhance the absorption efficiency while reducing the electrical cross-section of the device. For UV photodetectors, GaN NWs present the advantage of being mechanically and chemically robust. Within a GaN NW, it is possible to implement axial and radial (core-shell) heterostructures with AlGaN or InGaN materials, and the incorporation of heterostructures in GaN NW photodetectors opens interesting opportunities of performance improvement \cite{rigutti_ultraviolet_2010}. In particular, the internal electric field in polar GaN/AlN nanowire heterostructures can be used to modulate the spectral response, and even render it bias-dependent.

We performed several studies on electrically contacted GaN/AlN NW heterostructures operated as UV photodetectors. The observation of the exact same NW with TEM that is previously characterized by opto-electrical measurements is very powerful to guide modelling of the strain and band profile of the structure. For most studies, we used aberration-corrected high-angle annular dark field (HAADF) STEM, as a qualitative chemical contrast is easily observed using this technique.

In den Hertog \emph{et al.}\ \cite{den_hertog_correlation_2012}, we studied GaN NWs with an AlN insertion by correlated optoelectronic and aberration-corrected STEM characterization on the same single contacted NW, as summarized in figure~\ref{fig:hertog_nl}. Using annular bright field (ABF) and HAADF STEM, the NW growth axis was observed to be the N-polar $[000\bar{1}]$ direction [see figure~\ref{fig:hertog_nl}(g--i)]. The electrical transport characteristics of the NWs could be understood by the polarization-induced asymmetric potential profile combined with the presence of an AlN/GaN shell around the GaN wire base. We observed a higher forward current for increasing GaN outer shell thickness as demonstrated in figure~\ref{fig:hertog_nl}(a--f,j), which supports the hypothesis that the conduction takes place via the surface pathway created by the GaN shell. The AlN layer blocks the electron flow through the GaN core, confining the current to the radial GaN shell, close to the NW sidewalls, which increases the sensitivity of the photocurrent to the environment and in particular to the presence of oxygen. Based on our experiments and literature \cite{lefebvre_2012_oxygen}, we inferred that the desorption of oxygen adatoms in vacuum leads to a reduction of the non-radiative surface trap density, increasing both dark current and photocurrent.

 

\begin{figure}[t]
\includegraphics[width=0.49\textwidth]{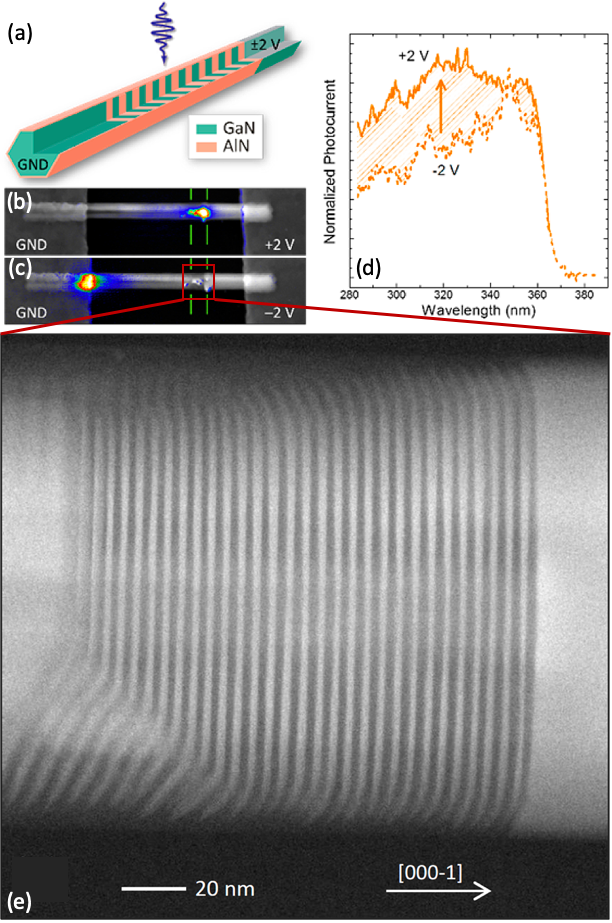}
\caption{\label{fig:Spies_nl} (a) Schematic description of the NW heterostructure. The voltage convention is indicated, with bias applied to the GaN cap (VB) whereas the GaN stem is grounded (GND). EBIC heat maps at (b) $-2$~V bias and (c) $+2$~V bias superimposed on an SEM image of the NW under study. The location of the GaN/AlN superlattice is outlined with green vertical lines. (d) Spectral response of the same single NW at bias voltages of $\pm2$~V. Data are corrected by the Xe-lamp emission spectrum taking the sublinear power dependence of the NW into account and normalized. Solid (dashed) lines correspond to positive (negative) bias. Shadowed areas outline the difference in the response between positive and negative bias. (e) HAADF STEM image of the superlattice for the studied NW. Adapted from reference \cite{spies_bias-controlled_2017} with permission from ACS.}
\end{figure}
 
In the work by Spies \emph{et al.}\ \cite{spies_bias-controlled_2017}, we presented a study of GaN-based single-NW
UV photodetectors with an embedded GaN/AlN superlattice, as summarized in figure~\ref{fig:Spies_nl}. The heterostructure dimensions and doping profile were designed in a way that the application of positive or negative bias leads to an enhancement of the collection of photogenerated carriers from the GaN/AlN superlattice or from the GaN base, respectively, as confirmed by electron beam-induced current (EBIC) measurements [figure~\ref{fig:Spies_nl}(b,c)]. The devices displayed enhanced response in the UV A (330--360~nm)/B (280--330~nm) spectral region under positive/negative bias, which could be understood by correlating the photocurrent measurements with STEM observations of the same single NW and semiclassical simulations of the strain and band structure in one and three dimensions. 

In general, the photocurrent of GaN NW photodetectors presents a sub-linear dependence on the illumination power 
\cite{gonzalez-posada_environmental_2013,lahnemann_uv_2016,lahnemann2017near,den_hertog_correlation_2012,spies_bias-controlled_2017}. However, Spies \emph{et al.}\ \cite{spies_effect_2018} demonstrated that such devices can behave as linear detectors, depending on the direction of the applied bias and the NW diameter. It was demonstrated that for diameters below a certain threshold, the complete depletion of the NW due to surface effects allows the fabrication of linear devices, if the NW heterostructure is properly designed.

\begin{figure*}
\includegraphics[width=0.99\textwidth]{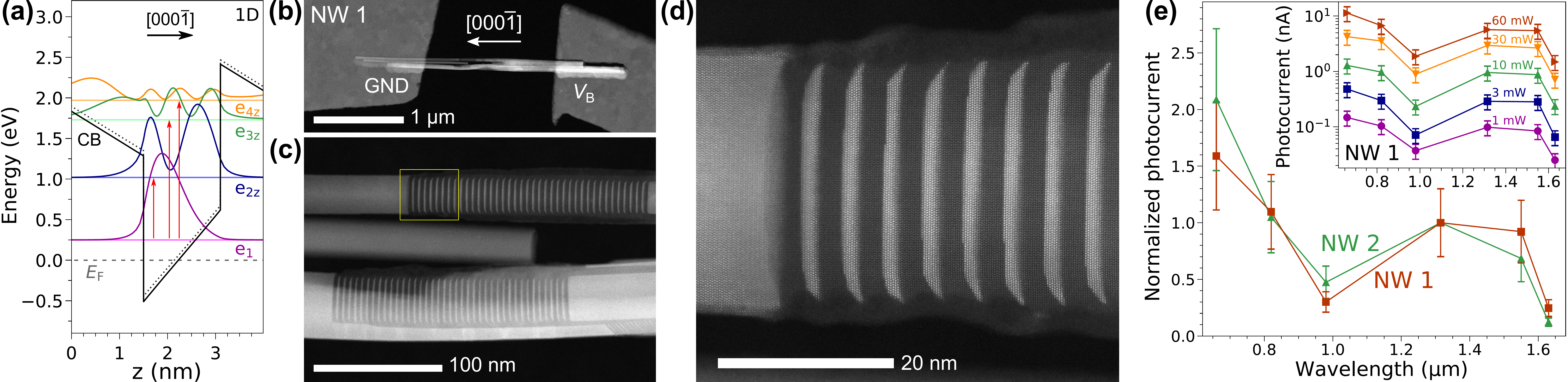}
\caption{\label{fig:laehnemann_nl}(a) One-dimensional \textbf{k}$\cdot$\textbf{p} calculations of the energy levels and wave functions in the quantum discs. The e$_1$--e$_{2z}$, e$_1$--e$_{3z}$ and e$_1$--e$_{4z}$ transitions are marked by the red arrows. The dotted profile shows the conduction band edge from three-dimensional calculations for comparison. (b) Overview HAADF STEM image of NW~1 and (c) detail of the GaN (bright)/AlN (dark) superlattice. (d) High-resolution HAADF STEM image of the region marked in panel (c) viewed along the $[21\bar{1}\bar{0}]$ direction. Both the growth direction and the contacting convention are labelled in panel (b). (e) Comparison of the normalized near-IR spectral photocurrent response for two NWs (NW~1 and NW~2) measured at 1~V bias. The response has been averaged over several different illumination powers. The inset shows the spectral response of NW~1 at different illumination levels on a semilogarithmic scale. The diameter of the laser spot at the membrane was always around 2~mm. The error bars account for the uncertainty in the calculation of the impinging irradiance due to the error in estimation of the spot size for the different laser diodes. Adapted from reference~\cite{lahnemann2017near} with permission from ACS.}
\end{figure*}

In the case of GaN-based UV photodetectors, we exploited the band-to-band absorption to generate a photocurrent. However, it is also possible to fabricate intersubband photodetectors, which rely on transitions between quantum-confined electron levels within the conduction band of quantum wells, as sketched in figure~\ref{fig:laehnemann_nl}(a). This principle enables IR photodetection even in wide band-gap semiconductors. We have demonstrated intersubband absorption in the near-IR telecommunication band using GaN/AlN superlattices embedded in single, contacted NWs. HAADF STEM images at different zoom levels in figure~\ref{fig:laehnemann_nl}(b)--(d) present such a single NW superlattice photodetector. The STEM images confirm the high quality and regularity of the heterostructures in a single NW, but show that in this particular case two NWs with well-separated superlattices are contacted in parallel. The dimensions of the quantum discs ($1.6\pm0.3$~nm) and barriers
($3.1\pm0.4$~nm) extracted from the STEM images allow us to determine the expected intersubband
transition energies via calculations of the band structure and energy levels in the NW superlattice, as presented in figure~\ref{fig:laehnemann_nl}(a). The spectral dependence of the IR photocurrent for this same NW, as well as another contacted NW from the same growth batch,  are represented in figure~\ref{fig:laehnemann_nl}(e). The increased response between 1.3 and 1.55~\textmu{}m and below 0.9~\textmu{}m correspond well to the calculated transition energies of 0.77, 1.4, and 1.71~eV for the e$_1$--e$_{2z}$, e$_1$--e$_{3z}$ and e$_1$--e$_{4z}$ transitions, respectively. 
In order to observe intersubband absorption, the wells in the nanowire are heavily doped. Therefore, the nanowire is not depleted and illumination with UV light results in a sublinear photoresponse. However, the photocurrent generated by IR radiation exhibits a linear dependence on the incident illumination power, which confirms that the intersubband process is insensitive to surface states. Additionally, the supplementary material of \cite{lahnemann2017near} shows that the coalescence of several NWs and presence of GaN shells around some of the superlattices can be correlated to the dark current levels of the contacted NWs.

\subsubsection{Heterostructured single nanowire devices: light emitters.}

Single semiconductor quantum dots are interesting objects for the implementation of photon sources \cite{shields_semiconductor_2007,mantynen_single-photon_2019}. In this field, III-nitride semiconductors offer significant advantages due to their large exciton binding energies and high band offsets, which enable room temperature operation of single photon sources at UV/visible emission wavelengths \cite{deshpande_electrically_2013,holmes_room-temperature_2014}. Here, an important challenge is to correlate electronic and structural properties in a single quantum dot, to reach a full understanding of such an object. This target is facilitated by the development of semiconducting NWs containing a single quantum dot, since single NWs can be easily dispersed on a Si$_3$N$_4$ membrane to correlate the emission properties of a contacted NW under electrical bias with its exact microstructure. Therefore, the slight variations in structural properties from NW to NW can be taken into account when studying varying optical and electrical behaviours.

\begin{figure}[t]
\includegraphics[width=0.49\textwidth]{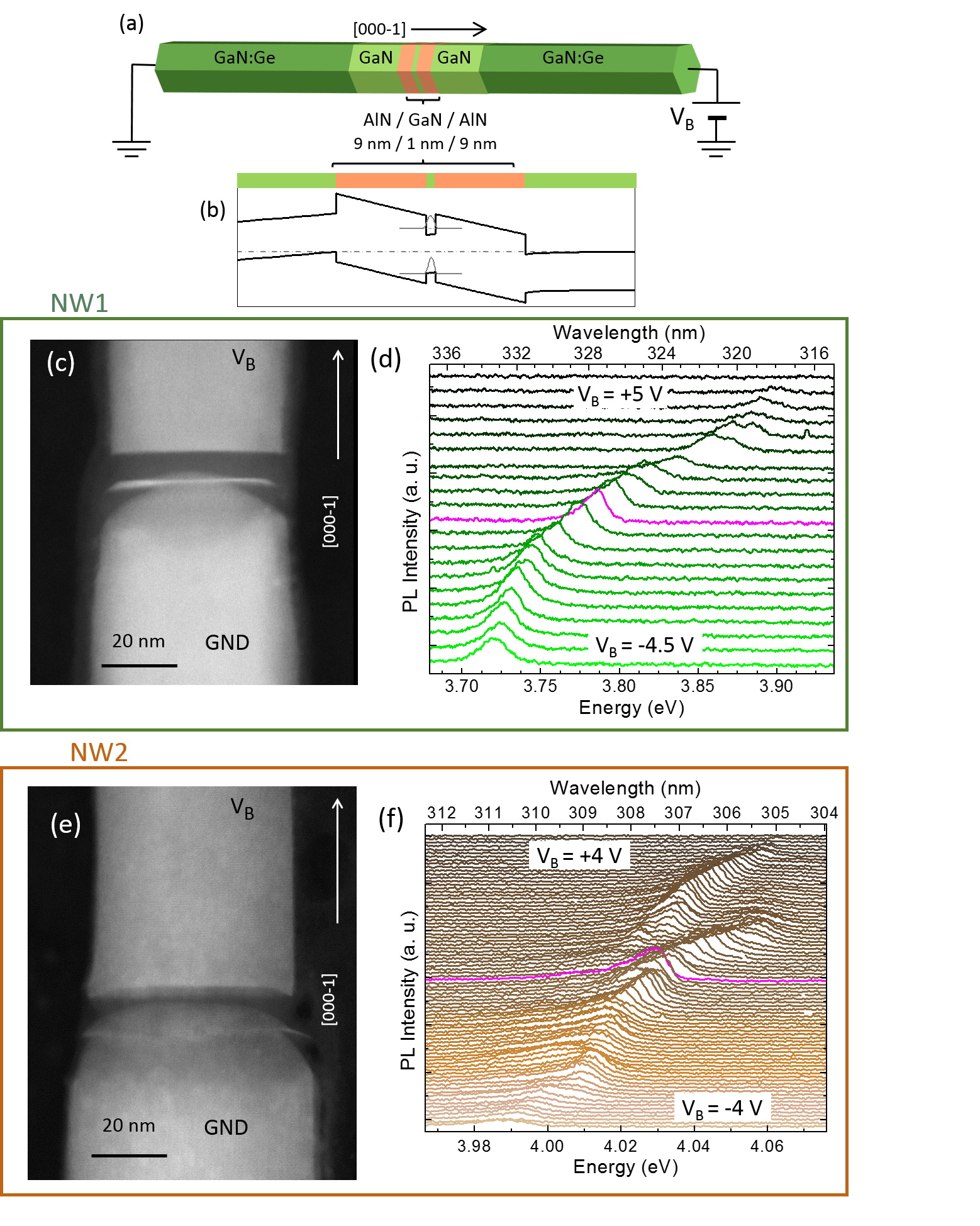}
\caption{\label{fig:spies_qd} (a) Simplified schematic of the NW structure with the AlN/GaN/AlN insertion in the centre. During the measurements, the NW stem is grounded and the bias is applied to the cap. (b) 1D nextnano3 simulations of the electric band structure of the nominal NW at zero bias. The squared wavefunctions of the electron and hole in the quantum disc are indicated. (c) HAADF-STEM micrograph of the heterostructure of NW1. \emph{I}-\emph{V} curve of NW1 in the dark and under UV illumination. (d) \textmu-PL spectra obtained applying bias from $-4.5$ to $+5$~V on NW1. The zero bias measurement is indicated in magenta. The spectra are given without normalization and shifted vertically for clarity. (e) \emph{I}-\emph{V} curve of NW2 in the dark and under UV illumination. (f) \textmu-PL spectra obtained applying bias from $-4$ to $+4$~V on NW2. The zero bias measurement is indicated in magenta. The spectra are given without normalization and shifted vertically for clarity. Adapted from reference \cite{spies2019correlated} with permission from ACS.}
\end{figure}

In reference~\cite{spies2019correlated}, STEM observations, photocurrent and \textmu{}-PL measurements under bias were performed on the same specimen, a GaN NW which contains a single AlN/GaN/AlN quantum disc structure, as sketched in figure~\ref{fig:spies_qd}(a). The corresponding band profile of the heterostructure is given in figure~\ref{fig:spies_qd}(b). 
More specifically, by applying an external bias, the spectral tunability of the emission of a single GaN quantum disc has been studied. The emission wavelength of a single quantum disc was observed to shift blue or red when the external electric field compensates or enhances the internal electric field generated by the spontaneous and piezoelectric polarization. A detailed study of two NW specimens is presented in figures~\ref{fig:spies_qd}(c)--(f). Without bias, these quantum discs emit at different wavelengths: 327.5~nm and 307.5~nm. Under external bias, the emission shows different spectral shifting rates of 20 and 12~meV/V, respectively. Theoretical calculations facilitated by the modelling of the exact heterostructure of each NW obtained by STEM imaging provide a good description of the experimental observations and the varying behaviours. When the bias-induced band bending is strong enough to favour tunnelling of the electron in the dot towards the stem or the cap, the spectral shift saturates and additional transitions associated to charged excitons can be observed. This depends on the exact barrier structure and the internal field in the quantum dot without applied bias.

\subsection{In-situ heating experiments}

\begin{figure}[t]
\includegraphics[width=0.49\textwidth]{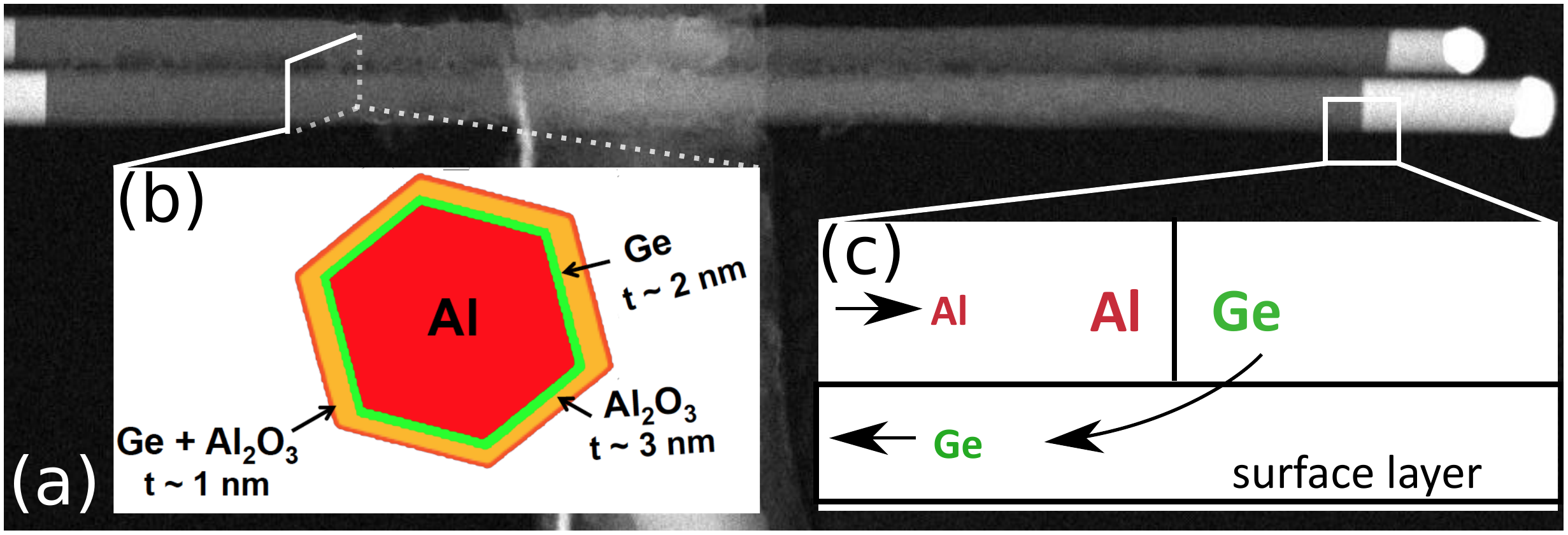}
\caption{\label{fig:joule-heating} (a) HAADF STEM image of two annealed Ge NWs side by side with Al contacts. A thermally-induced diffusion leads to an exchange between the Ge and Al atoms. (b) Schematic of the cross-section of the exchanged region that has a pure Al core, surrounded by a pure Ge shell and an Al$_2$O$_3$ shell enriched in Ge at the surface, as observed by model-based EDX analysis. (c) Schematic illustrating the diffusion process: Al is supplied to the reaction interface by self diffusion through the created Al core, while Ge diffuses into the Al contact by surface diffusion. Adapted from reference \cite{el2019situ}.}
\end{figure}

Local Joule heating using a metal strip on a semiconducting NW can be used to propagate a quasi-metallic phase into the NW, as first demonstrated by Mongillo \emph{et al.}\ \cite{mongillo_joule-assisted_2011} using electrical biasing in SEM. This is an interesting approach to obtain an atomically abrupt contact with low electrical resistance on NWs of group IV (Si and Ge). Using the electrical contacts on the NW for in-situ heating, we have studied the thermally-assisted solid state reaction in Al contacts on Ge and SiGe NWs \cite{el2019situ}, and Cu contacts on Ge NWs \cite{el_hajraoui_-situ_2019}. We also used our fabricated membrane-chips to study the Al propagation in Ge/Si core-shell NWs \cite{sistani_2018_monolithic}.

\begin{figure}[t]
\includegraphics[width=0.49\textwidth]{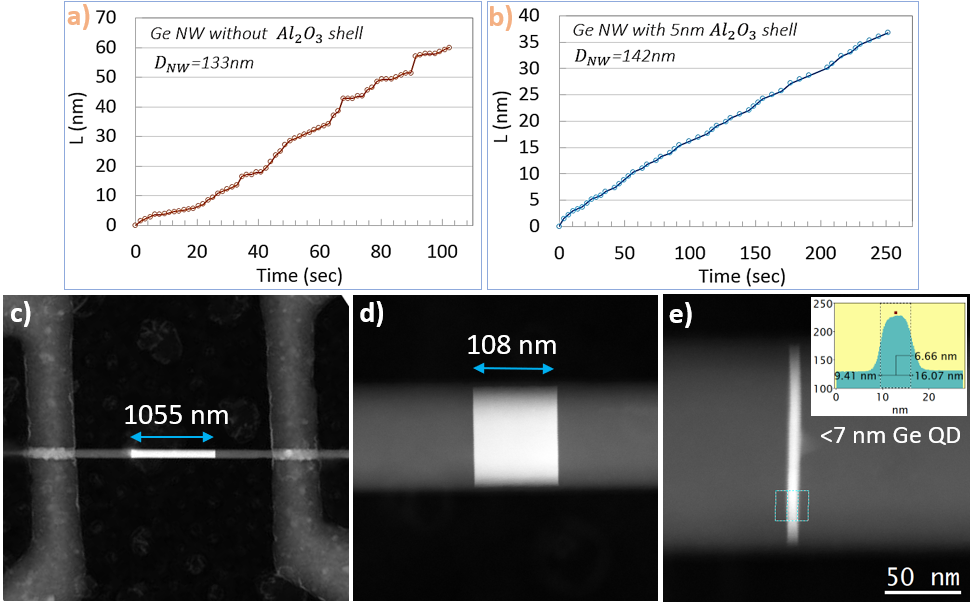}
\caption{\label{fig:Minh} (a-b) Plots of Al converted length $L$ as a function of time in the axial direction of the Ge NW without and with the protection of a 5 nm Al$_2$O$_3$ passivation shell, respectively. (c-e) HAADF STEM images of the Ge segment length at three different steps: (c) after RTA at 300 $^\circ$C during 20 s, (d) after a first Joule heating using the left heating electrode and (e) after a second Joule heating using the right electrode. Adapted from reference \cite{luong_2020} with permission from ACS.}
\end{figure}

Joule heating experiments were performed  in the TEM, using both a metal line patterned directly on the NW that can also be used as electrical contact, or using temperature calibrated heater chips from DENSsolutions. The TEM provides us both an ultimate spatial resolution of what happens during this reaction, as well as the possibility to use extensive TEM-based chemical characterization tools such as energy electron loss spectroscopy (EELS) or EDX prior to, during and at the end of the reaction. In figure~\ref{fig:joule-heating}, the Ge/Al NW-metal combination is shown. The HAADF STEM image of the reacted NW [figure~\ref{fig:joule-heating}(a)] shows an abrupt interface between the original Ge NW (bright contrast) and the converted region (darker contrast). Model-based EDX analysis combined with electron diffraction demonstrated that a mono-crystalline, pure Al NW was formed, surrounded by a shell of pure Ge and a shell of Al$_2$O$_3$ as illustrated in figure~\ref{fig:joule-heating}(b). Kinetic experiments revealed that the reaction rate is limited by a diffusion process, however no unambiguous influence of the NW diameter on the reaction rate was observed. Therefore, the reaction rate appears to be limited by self diffusion of Al through the created Al segment, and Ge diffuses into the Al contact by surface diffusion, as illustrated in figure~\ref{fig:joule-heating}(c). In the Cu/Ge NW couple, similar experiments demonstrated that the reaction rate is limited by surface diffusion of Ge. In this system, surface diffusion appears to occur both for Cu as well as Ge \cite{el_hajraoui_-situ_2019}. 
In the Al/Ge NW couple, we observed that the exchange reaction in large diameter ($>$100 nm) NWs does not proceed smoothly, but in nm sized jumps (see figure~\ref{fig:Minh}(a), where $L$ is the length of the converted region) \cite{luong_2020}.
In an effort to control the diffusion rate with atomic level precision, as grown Ge NWs were dipped in hidriodic acid (HI) for 5 s and were immediately passivated by a 5 nm Al$_2$O$_3$ shell using atomic layer deposition (ALD). It is interesting that a smooth and well-controlled propagation can be obtained in passivated Ge NWs, while the propagation in un-passivated NWs exhibits step-wise diffusion behaviours [see figure~\ref{fig:Minh}(a-b)]. The smoother advancement of the reaction interface in Al$_2$O$_3$ passivated NWs is attributed to an improvement of the surface quality compared to that of as-grown NWs, that may contain surface defects related to the native oxide. The discontinuous diffusion behaviour may then be explained by the trapping and de-trapping of the reaction interface at these defects. A reliable procedure for the fabrication of sub-10 nm Ge quantum dots, combing both ex-situ heating via rapid thermal annealing (RTA) and in-situ direct Joule heating technique, was proposed by Luong \emph{et al.}\ \cite{luong_2020}. Figure 12(c-e) shows HAADF STEM images of different Ge segment lengths produced at different steps of the heating procedure, finally demonstrating fabrication of a Ge quantum disk of 7 nm.

Advantages of our home fabricated membrane-chips with respect to commercial heater chips are the ease of fabricating many contacted NWs. However, the as-defined metal heater lines cannot be easily calibrated in temperature, while the commercial heater chips are well calibrated in temperature. A solution to this problem is to use a TEM-based method to measure the temperature \cite{Mecklenburg_2015_Nanoscale,lagos2018thermometry}, or to deposit small particles of a material with known and optimally chosen melting temperature on one side of the membrane-chip \cite{Bintlinger_2008_electron}.

\subsection{In-situ biasing experiments}

\begin{figure}[t]
\includegraphics[width=0.49\textwidth]{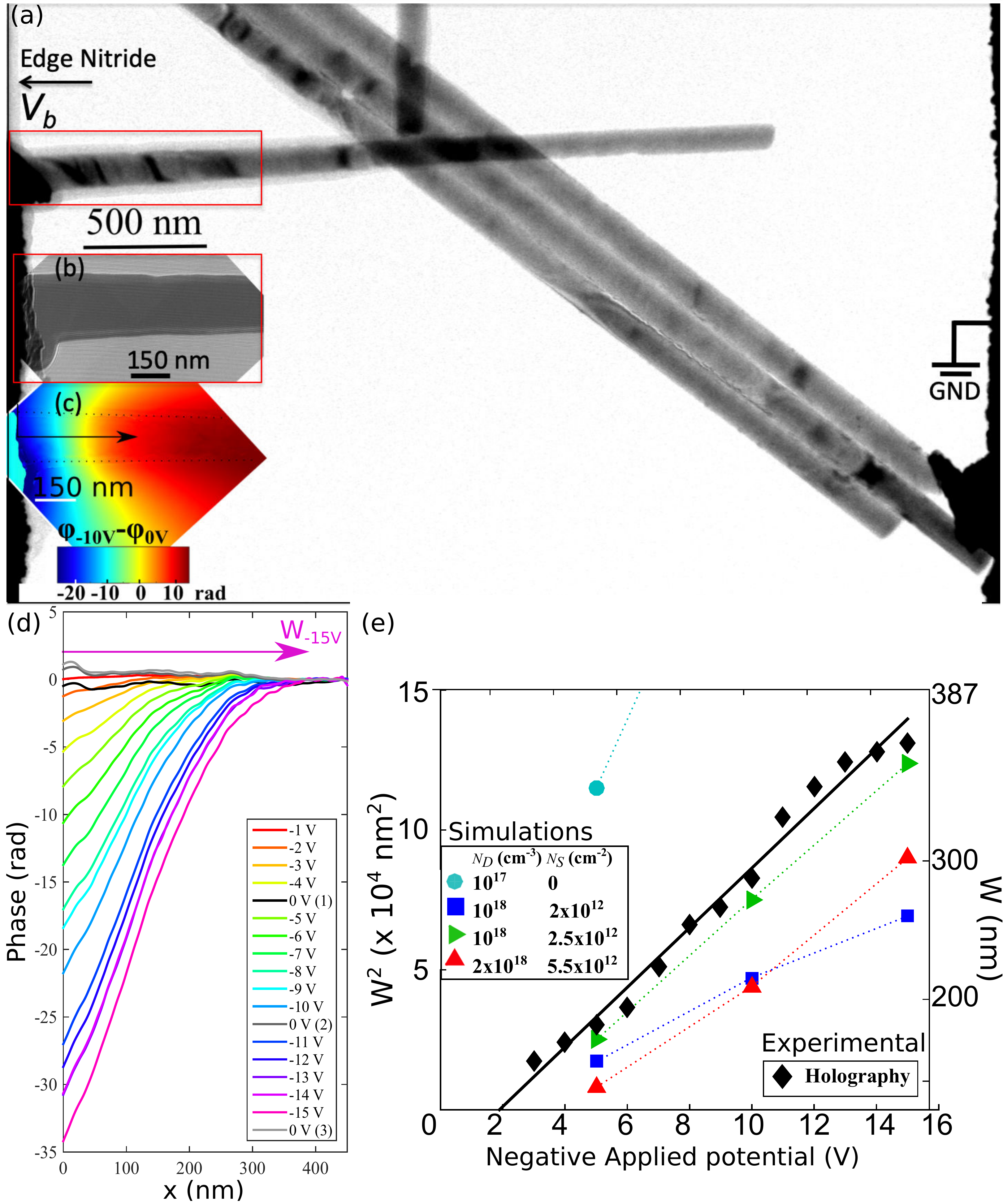}
\caption{\label{fig:hertog_bias}
(a) TEM image of a contacted, suspended ZnO NW. The reverse bias ($V_b$) was applied to the contact on the left side of the image. The NW is connected to the contact on the right side of the image (grounded) through other crossing NWs. (b) Zero-bias hologram obtained on the boxed region in (a). (c) Phase image with $-10$~V applied bias, after subtraction of the phase at zero bias. (d) Phase profiles obtained at the centre of the NW as a function of the reverse bias, obtained along the arrow shown in (c). The approximate depletion length at $V_b=-15$~V as obtained from these traces is indicated with a pink arrow. (e) 
Comparison of experimental depletion widths (black diamonds) obtained in (d) and fit with equation describing the depletion length in bulk semiconductors (solid line) with 3D calculations using the Nextnano3 software (symbols connected by dotted lines) including different doping $N_D$ and negative surface charge $N_S$ values for the NW. Adapted from reference \cite{den2017situ} with permission from IOP. }
\end{figure}

Quantitative characterization of electrically active dopants and surface charges in nano-objects is
challenging, since most characterization techniques using electrons \cite{voyles,wells_occurrence_1991,batson_simultaneous_1993}, ions \cite{zelsacherSims} or field ionization effects \cite{APTxBlavette,blavette_atom_1993,moutanabbir_colossal_2013} study the chemical presence of dopants, which are not necessarily electrically active. It was already shown that in-situ biasing in combination with off-axis electron holography can provide quantitative dopant characterization of electrically active dopants in a p-n junction in bulk samples \cite{twitchett2002}. The bias-dependent measurements provide an elegant way to separate the internal electric properties from other influences on the phase measured during electron holography. In \cite{den2017situ}, we reported a study using in-situ biasing in combination with off-axis electron holography to study the electrical properties of a Schottky contact on a ZnO NW. We performed cathodoluminescence and voltage contrast experiments on a contacted and biased suspended ZnO NW with a Schottky contact to measure the depletion length as a function of reverse bias. We compared these results with state-of-the-art off-axis electron holography in combination with electrical in-situ biasing on exactly the same NW, as shown in figure~\ref{fig:hertog_bias}. For off-axis electron holography experiments, it is preferred to have vacuum around the area of interest, so that the reference wave passing through vacuum is least perturbed. For this reason, the ZnO NW was suspended over a slit in the membrane-chip.

The extension of the depletion length under bias observed in techniques based on SEM is unusual as it follows a linear rather than square-root dependence, and is therefore difficult to model by bulk equations or finite element simulations. In contrast, the analysis of the axial depletion length observed by holography, see figure~\ref{fig:hertog_bias}(e), can be directly compared with 3D simulations.

This comparison of experiment and modelling permitted an estimation of a \emph{n}-type doping level of $1\times10^{18}$~cm$^{-3}$ and a negative sidewall surface charge of $2.5\times10^{12}$~cm$^{-2}$ for the NW, resulting in a radial surface depletion to a depth of 36~nm, in good agreement with values found in literature \cite{lord_factors_2013,shalish2004size}. We found excellent agreement between the simulated diameter of the conducting (non-depleted) core and the active thickness observed in the experimental data. By combining TEM holography experiments and finite element simulations of the NW electrostatics using the effective mass approximation, the bulk-like character of the NW core was revealed.

\section{Summary \& conclusions}

We have demonstrated a detailed fabrication process of silicon nitride membrane-chips compatible with both correlated ex-situ and in-situ opto-electrical experiments in TEM. 

We have shown that such custom-made membrane-chips can be used for the study of a variety of nanostructures by a broad range of correlated and in-situ characterization techniques. Our review highlights their potential for correlated microscopy studies, in-situ heating experiments, as well as in-situ electrical biasing experiments. For ex-situ studies, they can be used to identify the same nano-objects in subsequent measurements. Thereby, optical emission features can be attributed to specific structural defects, or the emission and absorption properties of a semiconductor heterostructure can be correlated to its precise structural properties, which increases the accuracy of theoretical modelling. 
In-situ Joule heating can be applied to observe solid state reactions such as the propagation of metal into a semiconductor NW. Finally, in-situ biasing during off-axis electron holography allows discriminating electronic effects from other influences on the phase, in order to extract information about doping levels and depletion regions.
Therefore, these experiments can give access to the charge, electric field and electrostatic potential distribution.
The advantage of our approach and the use of custom-fabricated membrane-chips is the compatibility with lithographic methods to establish electrical contacts with the NWs. Commercial membrane-chips are not optimised for lithography as typically no suitable markers for EBL are defined on the chip (both to localize the object or calibrate the EBL writing field), and these are sold per chip complicating the batch contacting of several NWs in the same EBL step. Thus, the main advantages of home made membrane-chips are:
\begin{itemize}
\item Batch contacting and easier handling during the
process.
\item A variety of marker structures to facilitate
contacting of nanowires and later characterization
on the same nanowires.
\item The membrane-chips can be entirely customized and adapted to the needs of the experiment, while it is difficult to commercialize a large variety of different chips. 
\end{itemize}

The main limitation of the home made chips is that to obtain a reliable temperature calibration, substantial additional development is needed.

The results presented here focus on the analysis of NW structures. However, our method can be used for different types of nano-objects, including 2D materials. Many more types of experiments can be realized on such membrane-chips. For example, micro-manipulators could be used to deterministically place the NW on a certain location or across a slit in the membrane in order to obtain a suspended and contacted NW or other object. Furthermore, contacts on the back side of the membrane-chip can be added, and used either as a back gate, or to influence the magnetization state of the sample. Moreover, while we have only shown experiments making use of direct currents for the biasing, also high frequency experiments could be realized on such a membrane-chip.



\ack
Financial support from the AGIR project CoPTon by the PEM P\^{o}le of Universit\'{e} de Grenoble-Alpes, the ANR JCJC COSMOS (ANR-12- 467 JS10-0002), the French CNRS and CEA is acknowledged. We acknowledge support from the Laboratoire d'excellence LANEF in Grenoble (ANR-10-LABX-51-01). This project has received funding from the European Research Council (ERC) under the European Union’s Horizon 2020 research and innovation programme (Grant Agreement 758385). We benefited from the access to the Nano characterization platform (PFNC) in CEA Minatec Grenoble in collaboration with the LEMMA group.

\bibliography{membranfabnotes_mdh}

\end{document}